\pdfoutput=1

\documentclass[prl,aps,twocolumn,showpacs,preprintnumbers,amssymb,nofootinbib,raggedbottom,10pt,longbibliography]{revtex4-2}

\usepackage{hyperref,graphicx,array,multirow,color,amsmath,mathrsfs,titlesec}
\titleformat{\section}{\centering\normalsize\normalfont\bf}{\thesection}{1em}{}

\usepackage[makeroom]{cancel}
\usepackage{amsmath,amssymb}
\usepackage{todonotes}
\usepackage{pgfplots}
\pgfplotsset{compat=1.15}
\usepackage{tikz}
\usetikzlibrary{calc,3d,fillbetween}
\def\uF{{u}}
\def\vF{{v}}
\def\wF{{w}}

\makeatletter\DeclareRobustCommand*{\bfseries}{\not@math@alphabet\bfseries\mathbf\fontseries\bfdefault\selectfont\boldmath}\makeatother

\begin{document}
\title{Folding Amplitudes into Form Factors: An Antipodal Duality}

\author{Lance~J.~Dixon$^{1}$, {\"O}mer G{\"u}rdo{\u{g}}an$^{2}$, Andrew~J.~McLeod$^{3,4,5}$ and Matthias Wilhelm$^{5}$}

\affiliation{$^1$ SLAC National Accelerator Laboratory,
Stanford University, Stanford, CA 94309, USA}

\affiliation{$^2$ School of Physics \& Astronomy, 
University of Southampton, Southampton, SO17 1BJ, UK}

\affiliation{$^3$ CERN, Theoretical Physics Department, 1211 Geneva 23, Switzerland}

\affiliation{$^4$ Mani L. Bhaumik Institute for Theoretical Physics, Department of Physics and Astronomy,
UCLA, Los Angeles, CA 90095, USA}

\affiliation{$^5$ Niels Bohr International Academy, Niels Bohr Institute,
  Blegdamsvej 17, 2100 Copenhagen \O{}, Denmark}

\preprint{SLAC-PUB-17637}

\begin{abstract}
We observe that the three-gluon form factor of the chiral part of the stress-tensor multiplet in planar $\mathcal{N}=4$ super-Yang-Mills theory is dual to the six-gluon MHV amplitude on its parity-preserving surface. Up to a simple variable substitution, the map between these two quantities is given by the antipode operation defined on polylogarithms (as part of their Hopf algebra structure), which acts at symbol level by reversing the order of letters in each term. We provide evidence for this duality through seven loops.  
\end{abstract}
\maketitle

\section{Introduction}\label{introduction_section}

In the study of quantum field theory, we occasionally encounter dualities, or relations between seemingly unrelated quantities.
One such example is the duality between scattering amplitudes and closed light-like polygonal Wilson loops in planar maximally supersymmetric Yang-Mills ($\mathcal{N}=4$ SYM) theory \cite{Alday:2007hr,Drummond:2007aua,Brandhuber:2007yx,Drummond:2007cf,Drummond:2007au,Alday:2008yw,Adamo:2011pv}, and its extension to a triality relating both quantities to a particular kinematic limit of correlation functions of the stress tensor supermultiplet~\cite{Alday:2010zy,Eden:2010zz,Eden:2010ce,Eden:2011yp,Eden:2011ku}.  
These types of relations provide us with valuable new perspectives on physical quantities, and at times reveal deep and novel types of mathematical structure.
In this letter, we present a new weak-weak duality between the maximally-helicity-violating (MHV) three-gluon form factor of the chiral part of the stress tensor supermultiplet in planar $\mathcal{N}=4$ SYM theory, and a kinematic limit of the six-gluon MHV amplitude in the same theory. This duality holds order-by-order in the {}'t~Hooft coupling $g^2=\frac{\lambda}{16\pi^2}$~\cite{tHooft:1973alw}.

A great deal is known about both the three-gluon form factor and the six-gluon amplitude. Their infrared structure can be understood to all orders in terms of the Bern-Dixon-Smirnov (BDS) ansatz~\cite{Bern:2005iz,Brandhuber:2012vm}, and each is known to be dual to a polygonal Wilson loop (which, in the case of the form factor, is periodic)~\cite{Alday:2007he,Alday:2007hr,Drummond:2007aua,Brandhuber:2007yx,Drummond:2007cf,Drummond:2007au,Alday:2008yw,Maldacena:2010kp,Adamo:2011pv,Brandhuber:2010ad,Ben-Israel:2018ckc,Bianchi:2018rrj}. Moreover, integrability techniques have been leveraged to develop an operator product expansion (OPE) around the near-collinear limit of each quantity~\cite{Basso:2013vsa, Basso:2013aha,Basso:2014koa,Basso:2014nra,Basso:2014hfa,Basso:2015rta,Basso:2015uxa,Belitsky:2014sla,Belitsky:2014lta,Belitsky:2016vyq,Sever:2020jjx,Sever:2021nsq,Sever:2021xga}, which has provided useful boundary data for bootstrap approaches, by means of which the amplitude has been computed through seven loops~\cite{Caron-Huot:2011dec,Dixon:2013eka,Dixon:2014iba,Dixon:2015iva,Dixon:2016apl,Caron-Huot:2016owq,Caron-Huot:2019bsq,Caron-Huot:2019vjl,Caron-Huot:2020bkp} and the form factor through eight loops~\cite{Dixon:2020bbt,ToAppearEightLoop}.  As will prove important below, both quantities are expressible in terms of multiple polylogarithms. This class of functions comes equipped with a coaction and an associated antipode (or coinverse)~\cite{Gonch3,Gonch2,Goncharov:2010jf,Brown:2011ik,Duhr:2011zq,Duhr:2012fh,2011arXiv1101.4497D}; see \cite{Duhr:2014woa} for a review. 

In this letter, we show that the antipode relates the three-gluon form factor to the six-gluon amplitude on its parity-preserving surface, up to a simple mapping between their respective kinematic variables. While it is surprising for any direct relation between these quantities to exist, the fact that they are related by the antipode map---which has no clear physical interpretation---is doubly bizarre.\footnote{At two loops, a different relation between the three-gluon form factor and six-gluon amplitude was observed that did not involve the antipode~\cite{Brandhuber:2012vm}. However, that relation does not hold beyond two loops \cite{Dixon:2020bbt}. An antipodal relation between one-loop integrals and Aomoto polylogarithms was found in \cite{Arkani-Hamed:2017ahv}.}

In the remainder of this letter, we provide the full statement of this relation, and present evidence that supports it through seven loops. We also discuss how this duality is consistent with many of the known analytic features of the six-gluon amplitude and three-gluon form factor, and draw out the implications this relation has for the analytic properties of the form factor, and for bootstrapping these quantities at higher loops.

\section{The Duality}
\label{sec:relation}

Let us first define the specific quantities that enter the amplitude/form factor duality that we find. On the amplitude side, we consider the BDS-like and cosmically-normalized six-point MHV amplitude $A_6$~\cite{Caron-Huot:2019bsq,Basso:2020xts}, which is defined by dividing the full amplitude $\mathcal{A}^{\text{MHV}}_6$ by the BDS-like ansatz $\mathcal{A}^{\text{BDS-like}}_6$ \cite{Alday:2009dv,Yang:2010as} and a transcendental function $\hat{\rho}$ \cite{Basso:2020xts}, which is independent of the kinematics but can be perturbatively expanded in the coupling $g^2$:
\begin{equation}
\label{eq: BDS-like cosmic A6}
\mathcal{A}^{\text{MHV}}_6 = \mathcal{A}^{\text{BDS-like}}_6 \times \hat{\rho} \times A_6 \, .
\end{equation}
$A_6$ is a finite polylogarithmic function%
\footnote{In particular, $A_6$ is the function referred to as $\mathcal{E}_{\text{cosmic}}$ in~\cite{Basso:2020xts}, which differs from the $\mathcal{E}$ function defined in~\cite{Caron-Huot:2019vjl} by a different choice for the function $\rho$, which we denote here by $\hat{\rho}$.}
of the three dual-conformally-invariant cross ratios
\begin{equation}
 \hat{u}=\frac{s_{12}s_{45}}{s_{123}s_{345}}\, ,\quad
 \hat{v}=\frac{s_{23}s_{56}}{s_{234}s_{123}}\, ,\quad
 \hat{w}=\frac{s_{34}s_{61}}{s_{345}s_{234}}\, , \label{cross-ratios}
\end{equation}
where $s_{i\dots k} = (p_i + \cdots +p_k)^2$ are planar Mandelstam invariants. In addition to rational functions of these cross ratios, $A_6$ depends on the square root of the six-point Gram determinant, which takes the form
\begin{equation}
\Delta=(1{-}\hat{u}{-}\hat{v}{-}\hat{w})^2-4 \hat{u} \hat{v} \hat{w} \, .
\end{equation}
Spacetime parity acts on the amplitude through the exchange $\sqrt{\Delta} \rightarrow -  \sqrt{\Delta}$.

\def\colorIvertical{red}
\def\colorIIIavertical{green}
\def\colorIIIbvertical{blue}
\def\colorIIIcvertical{orange}
\def\colorIhorizontal{violet}
\def\colorIIIahorizontal{teal}\def\colorIIIbhorizontal{cyan}\def\colorIIIchorizontal{yellow}
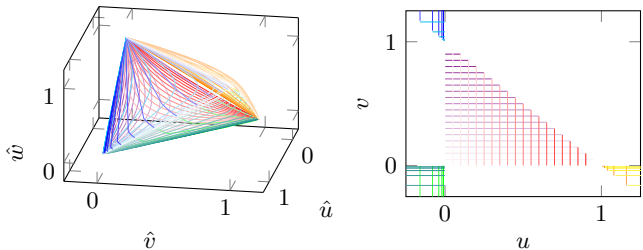
\begin{figure}[t]
   \begin{tikzpicture}[line width=.4pt,
   ]
   \begin{axis}[width=47mm,
      clip=true,
        xlabel=$\hat u$, ylabel=$\hat v$, zlabel=$\hat w$,
       xmin=-0.25,xmax=1.25,ymin=-0.25,ymax=1.25,zmin=-0.25,zmax=1.25,
      view = {100}{30},]
 
  \foreach [evaluate=\yi as \ycolor using \yi*100] \yi in {0.05,0.1,...,0.95} 
  {\edef\temp{\noexpand\addplot3 [domain=( +0.001):(1-\yi-0.001), samples=30,samples y=1,
               name path=xline,
               color=\colorIhorizontal!\ycolor, very thin,
           ]
            ({(\yi*(1-x-\yi)/((1-\yi)*(x+\yi)))^(1)} ,
               {(x*(1-x-\yi)/((1-x)*(x+\yi)))^(1)} ,
               {(x*\yi/((1-\yi)*(1-x)))^(1)} );}   
               \temp              
               } 
               
               \foreach [evaluate=\x as \xcolor using \x*100] \x in {0.05,0.1,...,0.95} 
  {\edef\temp{\noexpand\addplot3 [domain=( +0.001):(1-\x-0.001), samples=30,samples y=1,
               name path=xline,
               color=\colorIvertical!\xcolor, very thin,
           ]
            ({(x*(1-\x-x)/((1-x)*(\x+x)))^(1)} ,
               {(\x*(1-\x-x)/((1-\x)*(\x +x)))^(1)} ,
               {(\x*x/((1-x)*(1-\x)))^(1)} );}   
               \temp              
               }   
       \foreach 
      [evaluate=\i as \xcolor using (14-\i)/14*100] 
       [evaluate=\i as \x using 0.01*((2)^\i)] \i in {0,1,2,...,13} 
  {\edef\temp{\noexpand\addplot3 [domain=( +0.01):(100), samples=30,samples y=1,
               name path=xline,
               color=\colorIIIavertical!\xcolor
               , very thin,
           ]
            ({(-x*(1+\x +x)/((1+x)*(-\x -x)))^(1)} ,
               {(-\x*(1+\x +x)/((1+\x)*(-\x -x)))^(1)} ,
               {(\x*x/((1+x)*(1+\x)))^(1)} );}   
               \temp              
               }   
                    \foreach 
      [evaluate=\i as \xcolor using (14-\i)/14*100] 
       [evaluate=\i as \y using 0.01*((2)^\i)] \i in {0,1,2,...,13} 
  {\edef\temp{\noexpand\addplot3 [domain=( +0.01):(100), samples=30,samples y=1,
               name path=xline,
               color=\colorIIIahorizontal!\xcolor
               , very thin,
           ]
            ({(-\y*(1+x+\y)/((1+\y)*(-x -\y)))^(1)} ,
               {(-x*(1+x+\y)/((1+x)*(-x -\y)))^(1)} ,
               {(\y*x/((1+x)*(1+\y)))^(1)} );}   
               \temp              
               }          
               \foreach 
      [evaluate=\i as \xcolor using (14-\i)/14*100] 
       [evaluate=\i as \y using 0.01*((2)^\i)] \i in {0,1,2,...,13} 
  {\edef\temp{\noexpand\addplot3 [domain=( 1+\y+0.01):(100), samples=30,samples y=1,
               name path=xline,
               color=\colorIIIchorizontal!\xcolor
               , very thin,
           ]
            ({(-\y*(1-x+\y)/((1+\y)*(x -\y)))^(1)} ,
               {(x*(1-x+\y)/((1-x)*(x -\y)))^(1)} ,
               {(-\y*x/((1-x)*(1+\y)))^(1)} );}   
               \temp              
               }         
                \foreach 
       [evaluate=\i as \xcolor using (14-\i)/14*100] 
        [evaluate=\i as \y using 1+0.01*((2)^\i)] \i in {0,1,2,...,13} 
   {\edef\temp{\noexpand\addplot3 [domain=(0.001):(\y-1-0.001), samples=30,samples y=1,
                name path=xline,
                color=\colorIIIcvertical!\xcolor
                , very thin,
            ]   
             ({(-x*(1-\y +x)/((1+x)*(\y -x)))^(1)} ,
                {(\y*(1-\y +x)/((1-\y)*(\y -x)))^(1)} ,
                {(-\y*x/((1-\y)*(1+x)))^(1)} );}   
                \temp              
                }        
               \foreach 
      [evaluate=\i as \xcolor using (14-\i)/14*100] 
       [evaluate=\i as \y using 1+0.01*((2)^\i)] \i in {0,1,2,...,13} 
  {\edef\temp{\noexpand\addplot3 [domain=(0.001):(\y-1), samples=30,samples y=1,
               name path=xline,
               color=\colorIIIbhorizontal!\xcolor
               , very thin,
           ]
            ({(\y*(1+x-\y)/((1-\y)*(-x +\y)))^(1)} ,
               {(-x*(1+x-\y)/((1+x)*(-x +\y)))^(1)} ,
               {(-\y*x/((1+x)*(1-\y)))^(1)} );}   
               \temp              
               }        
               \foreach 
      [evaluate=\i as \xcolor using (14-\i)/14*100] 
       [evaluate=\i as \y using 0.01*((2)^\i)] \i in {0,1,2,...,13} 
  {\edef\temp{\noexpand\addplot3 [domain=(1+\y+0.01):(100), samples=30,samples y=1,
               name path=xline,
               color=\colorIIIbvertical!\xcolor
               , very thin,
           ]
            ({(x*(1+\y-x)/((1-x)*(-\y +x)))^(1)} ,
               {(-\y*(1+\y-x)/((1+\y)*(-\y +x)))^(1)} ,
               {(-\y*x/((1+\y)*(1-x)))^(1)} );}   
               \temp              
               } 
   \end{axis}
   \end{tikzpicture} 
\begin{tikzpicture}[line width=.4pt
]
   \begin{axis}[width=47mm,
      clip=true,xtick={0,1},ytick={0,1},
        xlabel=$u$, ylabel=$v$,
       xmin=-0.25,xmax=1.25,ymin=-0.25,ymax=1.25,]
 
  \foreach [evaluate=\yi as \ycolor using \yi*100] \yi in {0.05,0.1,...,0.95} 
  {\edef\temp{\noexpand\addplot [domain=( +0.001):(1-\yi-0.001), samples=30,samples y=1,
               name path=xline,
               color=\colorIhorizontal!\ycolor, very thin,
           ]
            ({x},{\yi} );}   
               \temp              
               } 
               
               \foreach [evaluate=\x as \xcolor using \x*100] \x in {0.05,0.1,...,0.95} 
  {\edef\temp{\noexpand\addplot [domain=( +0.001):(1-\x-0.001), samples=30,samples y=1,
               name path=xline,
               color=\colorIvertical!\xcolor, very thin,
           ]
            ({\x},{x} );}   
               \temp              
               }   
      \foreach 
      [evaluate=\i as \xcolor using (14-\i)/14*100] 
       [evaluate=\i as \x using 0.01*((2)^\i)] \i in {0,1,2,...,13}
  {\edef\temp{\noexpand\addplot [domain=( +0.001):(0.25), samples=30,samples y=1,
               name path=xline,
               color=\colorIIIavertical!\xcolor, very thin,
           ]
            ({-\x},{-x});}   
               \temp              
               }    \foreach 
      [evaluate=\i as \xcolor using (14-\i)/14*100] 
       [evaluate=\i as \x using 0.01*((2)^\i)] \i in {0,1,2,...,13}
  {\edef\temp{\noexpand\addplot [domain=( +0.001):(0.25), samples=30,samples y=1,
               name path=xline,
               color=\colorIIIahorizontal!\xcolor, very thin,
           ]
            ({-x},{-\x});}   
               \temp              
               }   
               \foreach 
      [evaluate=\i as \xcolor using (14-\i)/14*100] 
       [evaluate=\i as \y using 0.01*((2)^\i)] \i in {0,1,2,...,4} 
  {\edef\temp{\noexpand\addplot [domain=( 1+\y+0.01):(10), samples=30,samples y=1,
               name path=xline,
               color=\colorIIIbvertical!\xcolor
               , very thin,
           ]
            ({-\y},{x});}   
               \temp              
               }   
                \foreach 
       [evaluate=\i as \xcolor using (14-\i)/14*100] 
        [evaluate=\i as \y using 0.01*((2)^\i)] \i in {0,1,2,...,4} 
   {\edef\temp{\noexpand\addplot [domain=( 0.001):(\y), samples=30,samples y=1,
                name path=xline,
                color=\colorIIIbhorizontal!\xcolor
                , very thin,
            ]
             ({-x},{1+\y});}   
                \temp              
                }         
               \foreach 
      [evaluate=\i as \xcolor using (14-\i)/14*100] 
       [evaluate=\i as \y using 1+0.01*((2)^\i)] \i in {0,1,2,...,4} 
  {\edef\temp{\noexpand\addplot [domain=(0.001):(\y-1), samples=30,samples y=1,
               name path=xline,
               color=\colorIIIcvertical!\xcolor
               , very thin,
           ]
            ({\y} ,{-x} );}   
               \temp              
               }     
               \foreach 
      [evaluate=\i as \xcolor using (14-\i)/14*100] 
       [evaluate=\i as \y using 0.01*((2)^\i)] \i in {0,1,2,...,4} 
  {\edef\temp{\noexpand\addplot [domain=(1+\y+0.01):(10), samples=30,samples y=1,
               name path=xline,
               color=\colorIIIchorizontal!\xcolor
               , very thin,
           ]
            ({x} ,{-\y} );} 
               \temp              
               }   
   \end{axis}
   \end{tikzpicture} 
   \caption{Correspondence between various lines in the two-parameter three-point form factor space (right) and their images in the three-parameter six-point amplitude kinematic space (left) under the map \eqref{eq:kinematics_map_1}--\eqref{eq:kinematics_map_3}. The white regions in the right plot map to points outside of the region $0 < \hat{u}, \hat{v}, \hat{w} < 1$ in the left plot.
   }
   \label{fig:kinematics_map_1}
   \end{figure}

Similarly, we consider a BDS-like and cosmically-normalized version of the three-point MHV form factor $\mathcal{F}_3^{\text{MHV}}$,
\begin{equation}
\mathcal{F}^{\text{MHV}}_3 = \mathcal{F}^{\text{BDS-like}}_3 \times \rho \times F_3 \, ,
\end{equation}
where $\mathcal{F}^{\text{BDS-like}}_3$ was defined in \cite{Dixon:2020bbt} and $\rho$ is related to $\smash{\hat\rho}$ via the cusp anomalous dimension $\Gamma_{\text{cusp}}$: $\smash{\rho = \hat\rho \times\exp( - \zeta_2 \Gamma_{\text{cusp}}/2)}$.
$F_3$ is a finite polylogarithmic function%
\footnote{The function $\rho \times F_3$ was referred to as $\mathcal{E}$ in~\cite{Dixon:2020bbt}.}
 that depends on three ratios of Mandelstam invariants, which are usually chosen to be
\begin{equation}
 \uF=\frac{s_{12}}{s_{123}} \, ,\quad
 \vF=\frac{s_{23}}{s_{123}} \, ,\quad
 \wF=\frac{s_{13}}{s_{123}} \, . \label{ff_variables}
\end{equation}
We make use of all three variables in order to make manifest the (dihedral) symmetry of the form factor, but only two of these variables are independent due to momentum conservation, which implies that $\uF + \vF + \wF = 1$. 

The antipodal duality that we find between these quantities can be expressed as
\begin{equation}
F_3^{(L)}(u,v,w) = S \left( A_6^{(L)} (\hat{u},\hat{v},\hat{w})\right) \Big|_{\hat{u}_i \to \hat{u}_i(u,v,w)} \, , \label{ff_amp_relation}
\end{equation}
where $F_3^{(L)}$ and $A_6^{(L)}$ denote the $\mathcal{O}(g^{2L})$ contributions to $F_3$ and $A_6$, $S$ is the antipode map, and
\begin{align}
\hat{u}_1 = \hat{u}(u,v,w) &= \frac{\vF \wF}{(1-\vF)(1-\wF)} \, , \label{eq:kinematics_map_1} \\
\hat{u}_2 = \hat{v}(u,v,w) &= \frac{\uF \wF}{(1-\uF)(1-\wF)} \, , \label{eq:kinematics_map_2} \\
\hat{u}_3 = \hat{w}(u,v,w) &= \frac{\uF \vF}{(1-\uF)(1-\vF)}\, .  \label{eq:kinematics_map_3}
\end{align}
The antipode map is part of the larger Hopf algebra structure of multiple polylogarithms, which also contains the coproduct and symbol maps~\cite{Gonch2,Goncharov:2010jf,Brown:2011ik,Duhr:2011zq,Duhr:2012fh}. The symbol of a polylogarithmic function $G$ is recursively defined via its total differential as 
\begin{equation}
d G = \sum_{x\in \mathcal{L}} G^x\, d \ln x \quad \Rightarrow \quad \mathcal{S}(G) = \sum_{x\in \mathcal{L}} \mathcal{S}(G^x\,) \otimes x \, ,
\end{equation} 
where the set of logarithmic arguments $\mathcal{L}$ is referred to as the symbol alphabet, and each of the functions $G^x$ is also a polylogarithm.

At symbol level, the antipode map simply reverses the order of the letters in every word of the symbol (up to a sign) \cite{Gonch3,Brown:2013gia}:
\begin{equation}
\label{eq: antipode at symbol level}
S(x_1 \otimes x_2 \otimes \dots \otimes x_m ) = (-1)^m\ x_m \otimes \dots \otimes x_2 \otimes x_1 \, .
\end{equation}
However, we find that relation \eqref{ff_amp_relation} also holds\footnote{Strictly speaking, the antipode map only makes sense on de Rham periods, and as such is not defined on $i\pi$.} for terms involving transcendental constants, modulo contributions proportional to $i \pi$.

Due to the momentum conservation constraint on the form factor variables, the substitutions~\eqref{eq:kinematics_map_1}--\eqref{eq:kinematics_map_3} require the amplitude to be evaluated on a two-dimensional surface. In particular, the equation $u+v+w = 1$ gets mapped to the constraint that $\Delta =  0$. Since parity sends $\sqrt{\Delta}\to -\sqrt{\Delta}$, this is the surface on which the parity of the amplitude is preserved.

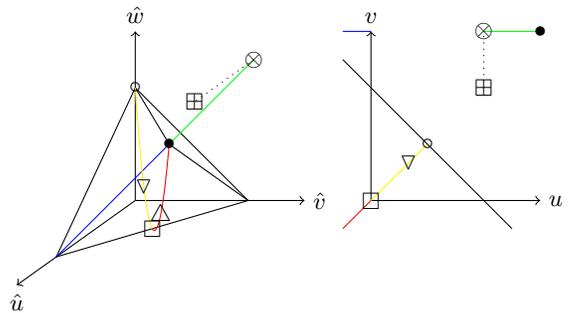
\begin{figure}[t]\begin{tikzpicture}[x  = {(-0.7cm,-0.5cm)},
                    y  = {(1cm,0cm)},
                    z  = {(0cm,1cm)},
                    scale = 1.5,baseline={(0,0,0)}]
\draw[->] (0,0,0) -- (1.5,0,0) node[left, below] {$\hat u$};
\draw[->] (0,0,0) -- (0,1.5,0) node[right] {$\hat v$};
\draw[->] (0,0,0) -- (0,0,1.5) node[above] {$\hat w$};
 \draw[] (1,0,0) -- (0,0,1);
 \draw[] (1,0,0) -- (0,1,0);
 \draw[] (0,0,1) -- (1,1,1);
 \draw[] (0,1,0) -- (1,1,1);
\draw[] (0,1,0) -- (0,0,1);
\draw[blue] (1,0,0) -- (1,1,1);
\draw[green] (1,1,1) -- (1,1.75,1.75);
\draw[dotted] (1,1.75,1.75) -- (1.75,1.75,1.75);
\draw[domain=0:1/2,yellow,smooth] plot ({\x},{\x},{(1-2*\x)^2});
\draw[domain=1/2:1,red,smooth] plot ({\x},{\x},{(1-2*\x)^2});
\node[] at (1/4,1/4,1/4){$\triangledown$};
\node[] at  (1/2,1/2,0){$\square$};
\node[] at  (1,1,1){$\bullet$};
\node[] at  (0,0,1){$\circ$};
\node[] at  (1,1.75,1.75){$\otimes$};
\node[] at  (3/4,3/4,1/4){$\triangle$};
\node[] at  (1.75,1.75,1.75){$\boxplus$};
\end{tikzpicture}
\begin{tikzpicture}[x  = {(1cm,0cm)},
                    y  = {(0cm,1cm)},
                    scale = 1.5,baseline={(0,0)}]
\draw[->] (0,0) -- (1.5,0) node[right] {$u$};
\draw[->] (0,0) -- (0,1.5) node[above] {$v$};
\draw[yellow] (0,0) -- (1/2,1/2);
\draw[red] (0,0) -- (-1/4,-1/4);
\draw[] (-0.25,1.25) -- (1.25,-0.25);
\draw[dotted] (1,1) -- (1,1.5);
\draw[green] (1,1.5) -- (1.5,1.5);
\draw[blue] (-0.25,1.5) -- (0,1.5);
\node[] at (1/3,1/3){$\triangledown$};
\node[] at  (0,0){$\square$};
\node[] at  (1.5,1.5){$\bullet$};
\node[] at  (1/2,1/2){$\circ$};
\node[] at  (1,1.5){$\otimes$};
\node[] at  (1,1){$\boxplus$};
\end{tikzpicture}
\caption{Schematic correspondence between various interesting points and lines in the two-parameter three-point form factor space (right) and their images in the three-parameter six-point amplitude kinematic space (left) under the map \eqref{eq:kinematics_map_1}--\eqref{eq:kinematics_map_3}. See Table~\ref{tab:mapping of points} for the coordinates of the marked points.
}
\label{fig:kinematics_map_2}
\end{figure}

\begin{table}[t]
\begin{tabular}{l|c|c|c}
 & $(\hat u,\hat v,\hat w)$ & $(u,v,w)$ & functions\\\hline
 $\triangledown$ & $(\frac{1}{4},\frac{1}{4},\frac{1}{4})$ & $(\frac{1}{3},\frac{1}{3},\frac{1}{3})$ & $\sqrt[6]{1}$\\[.03cm]
 $\square$ & $(\frac{1}{2},\frac{1}{2},0)$ & $(0,0,1)$ & $\text{Li}_2(\frac{1}{2})$ + logs\\
 $\bullet$ & $(1,1,1)$ & $\lim_{u\to\infty}(u,u,1{-}2u)$ & MZVs\\
 $\circ$ & $(0,0,1)$ &$(\frac{1}{2},\frac{1}{2},0)$ & MZVs + logs \\
 $\triangle$ & $(\frac{3}{4},\frac{3}{4},\frac{1}{4})$ & $(-1,-1,3)$ & $\sqrt[6]{1}$ \\
 $\boxplus$ & $(\infty,\infty,\infty)$ & $(1,1,{-}1)$ & alternating sums\\
 $\otimes$ & $\lim_{\hat{v}\to\infty}(1,\hat{v},\hat{v})$ & $\lim_{v\to\infty}(1,v,{-}v)$ & MZVs \\
 \begin{tikzpicture}
  \draw[green] (0,0) -- (1/5,0);
  \draw[blue] (1/5,0) -- (2/5,0);
 \end{tikzpicture}
& $(1,\hat{v},\hat{v})$ &$\lim_{v\to\infty}(u,v,1{-}u{-}v)$ & HPL$\{0,1\}$ \\
 \begin{tikzpicture}
  \draw[red] (0,0) -- (1/5,0);
  \draw[yellow] (1/5,0) -- (2/5,0);
 \end{tikzpicture}
& $(\hat{u},\hat{u},(1-2\hat{u})^2)$ &$(u,u,1-2u)$ & HPL$\{-1,0,1\}$
\end{tabular}
\caption{Kinematic points and lines as well as  their images under the map \eqref{eq:kinematics_map_1}--\eqref{eq:kinematics_map_3}.
The points $\triangledown$, $\square$, $\circ$, $\triangle$ and $\boxplus$ all lie on the line  
$(u,u,1-2u)$.
}
\label{tab:mapping of points}
\end{table}

We depict the mapping between these kinematical spaces in Figures~\ref{fig:kinematics_map_1} and~\ref{fig:kinematics_map_2}, and give the translation for various points and lines in Table \ref{tab:mapping of points}. 
In particular, Figure~\ref{fig:kinematics_map_2} and Table \ref{tab:mapping of points} show several interesting points and lines on which the multiple polylogarithms simplify to multiple zeta values (MZVs), alternating sums, cyclotomic zeta values~\cite{Ablinger:2011te} including $6^{\rm th}$ roots of unity \cite{HyperlogProcedures,Caron-Huot:2019bsq} (denoted in Table \ref{tab:mapping of points} by $\sqrt[6]{1}$), and harmonic polylogarithms (HPLs)~\cite{Remiddi:1999ew}. At two of the points the amplitude and form factor diverge logarithmically. Of the two HPL lines $(1,\hat{v},\hat{v})$ and $(\hat{u},\hat{u},(1-2\hat{u})^2)$, the former is simpler because the HPL index $-1$ does not appear. In the next section, we will check the duality at some of these points and lines. 

As can be seen in Figure~\ref{fig:kinematics_map_1}, the line $v=0$ is mapped to the point $(\hat{u},\hat{v},\hat{w})=(0,1,0)$, while the point $(u,v)=(1,0)$ is mapped to the line $\hat{u}=0$,\ \ $\hat{v}+\hat{w}=1$.
These relations imply that the duality exchanges soft and collinear limits.
This exchange is related to the simplicity of the map~\eqref{eq:kinematics_map_1}--\eqref{eq:kinematics_map_3} in the OPE parametrization:
\begin{equation}
\begin{gathered}
\hat{u}=\frac{1}{1+(\hat{T}+\hat{S}\hat{F})(\hat{T}+\hat{S}/\hat{F})} \,, \\
\hat{v}=\hat{u}\hat{w}\hat{S}^2/\hat{T}^2 \,, \qquad
\hat{w}=\frac{\hat{T}^2}{1 +  \hat{T}^2} \,,
\end{gathered}
\end{equation}
and 
\begin{equation}
\begin{gathered}
u = \frac{1}{1 + S^2 + T^2} \,, \qquad
v = \frac{T^2}{1 + T^2} \,, \\
w = \frac{1}{(1 + T^2)(1 + S^{-2} (1 + T^2))} \,,
\end{gathered}
\end{equation}
see \cite{Basso:2013vsa,Sever:2020jjx}.
Namely, 
\begin{equation}
\hat{T} = \frac{T}{S}\,, \qquad \hat{S} = \frac{1}{TS}\, ,
\end{equation}
while $\hat{F} = 1$ on the $\Delta = 0$ surface.

\section{Evidence}
\label{sec:evidence}

Since the six-point amplitude and three-point form factor have been computed through seven and eight loops~\cite{Caron-Huot:2019vjl,Dixon:2020bbt,ToAppearEightLoop}, respectively, we can provide evidence for relation~\eqref{ff_amp_relation} through seven loops. We first do so at symbol level, where the antipode map is given by \eqref{eq: antipode at symbol level}.

The alphabet of symbol letters of the six-point amplitude can be chosen to be 
\begin{equation}\label{eq:amplitude_letters}
\hat{\mathcal{L}} = \{\hat{a},\hat{b},\hat{c},\hat{d},\hat{e},\hat{f},\hat{y}_u,\hat{y}_v,\hat{y}_w \} \, ,
\end{equation}
where
\begin{align}
\hat{a} &= \frac{\hat{u}}{\hat{v}\hat{w}} \,, && \hat{b} = \frac{\hat{v}}{\hat{w}\hat{u}} \,,
&&  \hat{c} = \frac{\hat{w}}{\hat{u}\hat{v}} \,, \nonumber \\
\hat{d} &= \frac{1-\hat{u}}{\hat{u}} \,, && \hat{e} = \frac{1-\hat{v}}{\hat{v}} \,,
&&  \hat{f} = \frac{1-\hat{w}}{\hat{w}} \, ,
\end{align}
and the remaining variables invert under parity, for instance $\hat{y}_u\to1/\hat{y}_u$. Hence, $\hat{y}_u=\hat{y}_v=\hat{y}_w=1$ on the $\Delta=0$ surface.\footnote{Note that this notation differs from~\cite{Caron-Huot:2019vjl,Caron-Huot:2019bsq}, where $\smash{\hat{d}}$, $\smash{\hat{e}}$, and $\smash{\hat{f}}$ were referred to as $m_u$, $m_v$ and $m_w$.} An analogous alphabet can be chosen for the three-point form factor, namely
\begin{equation} \label{eq:abc_letters}
\mathcal{L} = \{a,b,c,d,e,f \} \, ,
\end{equation}
where
\begin{align}
a &= \frac{u}{vw} \,, &&  b = \frac{v}{wu} \,,
&&  c = \frac{w}{uv} \,, \nonumber \\
d &= \frac{1-u}{u} \,, && e = \frac{1-v}{v} \,,
&& f = \frac{1-w}{w} \,. \label{abcuvw}
\end{align}
Both $A_6|_{\Delta=0}$ and $F_3$ are invariant under the same dihedral group $D_3$, which is generated by 
\begin{equation}
\label{eq: dihedral cycle}
\text{cycle:\ } \{a,b,c,d,e,f\}\to\{b,c,a,e,f,d\}
\end{equation}
and 
\begin{equation}
\label{eq: dihedral reflection}
\text{flip:\ } \{a,b,c,d,e,f\}\to\{a,c,b,d,f,e\} \, ,
\end{equation}
and similarly in the hatted letters.

It is easy to check that the map~\eqref{eq:kinematics_map_1}--\eqref{eq:kinematics_map_3} acts by swapping
\begin{align}
\label{eq: letter identification}
 \sqrt{\hat{a}}     \Leftrightarrow    d \, , \qquad
\hat{d}            \Leftrightarrow   a \, , \qquad 
\end{align}
and all cyclically-related letters in the same way. Notably, this swaps the letters that appear in the first and last entries of the amplitude and form factor. That is, the letters that are allowed to appear in the first entry of the amplitude ($\smash{\hat{a}}$, $\smash{\hat{b}}$, and $\smash{\hat{c}}$) are mapped to the letters that appear in the last entry of the form factor ($d$, $e$, and $f$). Similarly, the first entries of the form factor ($a$, $b$ and $c$) are mapped to the last entries of the amplitude ($\smash{\hat{d}}$, $\smash{\hat{e}}$, and $\smash{\hat{f}}$). This combines with the reversal of symbol letters entailed by the antipode map in~\eqref{ff_amp_relation} to maintain the known first and last entry conditions~\cite{Gaiotto:2011dt,Caron-Huot:2011dec,Brandhuber:2012vm,Dixon:2020bbt} of each quantity.

We now check that relation~\eqref{ff_amp_relation} holds at symbol level. For example, at two loops, the symbols $\mathcal{S}$ of the amplitude and form factor each involve 12 terms, and are given by
\begin{align}
\mathcal{S}(A_6^{(2)}|_{\Delta=0})=\frac{1}{2}\,\hat{a}{\otimes}\hat{a}{\otimes}\hat{a}{\otimes}\hat{e}{+}\hat{a}{\otimes}\hat{e}{\otimes}\hat{e}{\otimes}\hat{e}{+}{\text{dihedral}}\,,\label{eq:amp_two_loop}\\
\mathcal{S}(F_3^{(2)})=2\,a{\otimes}a{\otimes}a{\otimes}e+4\,a{\otimes}e{\otimes}e{\otimes}e+\text{dihedral}\,,\label{eq:ff_two_loop}
\end{align}
where we sum over all dihedral images in $D_3$ generated by \eqref{eq: dihedral cycle} and \eqref{eq: dihedral reflection}. Due to the square root in~\eqref{eq: letter identification}, the two terms in~\eqref{eq:amp_two_loop} pick up factors of 8 and 2, which are precisely the numerical factors needed to match~\eqref{eq:ff_two_loop} when the rest of the transformation in~\eqref{ff_amp_relation} is applied. We have similarly checked that the duality holds at symbol level through seven loops, using the expressions provided in the ancillary files of~\cite{Caron-Huot:2019vjl} and~\cite{Dixon:2020bbt,ToAppearEightLoop}. Due to the fast growth of the number of terms in the symbol with the loop order (see Table \ref{tab: number of terms in the symbol}), this check quickly becomes quite involved, and extremely stringent.

\begin{table}[t]
\begin{tabular}{l|r}
$L$ & number of terms\\\hline
1 & 6 \\
2 & 12 \\
3 & 636 \\
4 & 11,208 \\
5 & 263,880 \\
6 & 4,916,466 \\
7 & 92,954,568 \\
8 & 1,671,656,292 
\end{tabular}
 \caption{Number of terms in the symbol of $F_3^{(L)}$ as a function of the loop order $L$.}
 \label{tab: number of terms in the symbol}
\end{table}

The antipode map is also defined at function level~\cite{Gonch3,Brown:2013gia} (see also~\cite{DelDuca:2016lad} for a discussion of the antipode in the physics literature), and as such we can also check the duality beyond the symbol. The simplest way to do this is to compare the functions at a single point, for instance at the point $\hat{u} = \hat{v} = \hat{w} = 1$, which maps to the $u,v\to\infty$ limit of the form factor space. At these points, both functions are real and are known to be expressible in terms of MZVs, which can conveniently be expressed in terms of the so-called $f$-alphabet \cite{Brown:2011ik,HyperlogProcedures}; see Table \ref{tab:mapping of points}. Since we do not know how to compute the antipode of $i\pi$, the first nontrivial constants appear at three loops. Through five loops, the amplitude evaluates to \cite[(A.3)--(A.5)]{Caron-Huot:2019bsq}
\begin{align}
A_6^{(3)}(1,1,1) &= 0 f_{3,3} + \mathcal{O}(\pi^2)  \, , \\
A_6^{(4)}(1,1,1) &= 120f_{3,5} + \mathcal{O}(\pi^2)  \, , \\
A_6^{(5)}(1,1,1) &=-2688  f_{3,7} - 1560  f_{5,5}  + \mathcal{O}(\pi^2) \, ,
\end{align}
while the form factor evaluates to \cite[(5.7)--(5.9)]{Dixon:2020bbt}
\begin{align}
F_3^{(3)}(\infty,\infty) &= 0 f_{3,3} + \mathcal{O}(\pi^2)  \, , \\
F_3^{(4)}(\infty,\infty) &= 120f_{5,3} + \mathcal{O}(\pi^2)  \, , \\
F_3^{(5)}(\infty,\infty) &= -2688  f_{7,3} - 1560  f_{5,5}  + \mathcal{O}(\pi^2)  \, .
\end{align}
Clearly, these values are related to each other by reversing the order of $f$-alphabet letters.\footnote{More generally, similar to its action on the symbol, the antipode reverses the $f$-alphabet letters and multiplies each term by $(-1)^m$, where $m$ is the number of letters.}
We provide further evidence for the duality at various points up to seven loops in an ancillary file.\footnote{This ancillary file also contains further details on the definitions of $\mathcal{F}^{\text{BDS-like}}_3$, $\mathcal{A}^{\text{BDS-like}}_6$, $\rho$, and $\hat\rho$.} 

We can also check the duality on the line where \mbox{$\hat{u}=1$} and $\hat{v}= \hat{w}$, where $A_6$ can be expressed in terms of HPLs with indices $0,1$ and argument $\hat{x} = 1 - 1/\hat{v}$. This line maps via \eqref{eq:kinematics_map_1}--\eqref{eq:kinematics_map_3} to the line where $v\to \infty$ (with $u$ fixed), where $F_3$ can be expressed in terms of the same space of functions, but with the arguments reinterpreted as $x = 1-1/u$, and $\hat x=1-x$. We have checked that these functions map to each other via relation~\eqref{ff_amp_relation} through seven loops, up to terms proportional to $\pi^2$.  We then use the duality to predict the eight-loop MHV amplitude on the line $(1,\hat{v},\hat{v})$, modulo $\pi^2$ terms, in a second ancillary file. Finally, we have also checked that relation~\eqref{ff_amp_relation} holds at the level of full functions of $u$ and $v$, up to three loops, finding a complete match up to terms proportional to $i\pi$. 
While we have not detailed here how the antipode acts in general on multiple polylogarithms, we note that it is conveniently implemented in the \textsc{Mathematica} package \texttt{PolyLogTools}~\cite{Duhr:2019tlz}.

It would be interesting to find an extension or deformation of relation \eqref{ff_amp_relation} that also relates the terms proportional to $i\pi$ on both sides of the duality. This is non-trivial, though, for two reasons. The first is that the antipode is not defined on $i\pi$. Second, there is a question of the appropriate Riemann sheets. On its physical sheet, the form factor is real when $0 < u,v,w < 1$ and complex elsewhere (except when one of these variables is taken to infinity). For the amplitude, with $\hat{u},\hat{v},\hat{w} > 0$, we could either be on the Euclidean sheet, or the $2 \to 4$ physical scattering sheet. 
Both are problematic: On the Euclidean sheet, the amplitude is real, while the form factor generically has imaginary parts proportional to $i\pi$ (but at least for $u,v,w<1$ both objects are real). On the $2 \to 4$ physical scattering sheet, the amplitude has imaginary parts, which blow up logarithmically as one approaches the $(1,\hat{v},\hat{v})$ self-crossing line \cite{Dixon:2016epj}, while the dual form factor's imaginary parts vanish there.

\section{Implications}
\label{sec:implications}

This duality has several interesting implications. The  six-point amplitude is known to obey a large set of extended Steinmann relations~\cite{Caron-Huot:2019bsq} (or cluster adjacency conditions~\cite{Drummond:2017ssj}), which tell us that certain pairs of letters never appear in adjacent entries of the symbol: 
\begin{equation}
\begin{gathered} 
\cancel{\dots \hat a \otimes \hat b \dots}\, ,  \\
\cancel{\dots \hat a \otimes \hat d \dots}\, , \quad  \cancel{\dots \hat d \otimes \hat a \dots}\, , 
\end{gathered}
\end{equation}
plus all dihedral images. Importantly, these conditions all remain nontrivial and distinct on the $\Delta = 0$ surface, and can be read either backwards or forwards.  As such, these constraints are preserved by the antipode, and can be translated directly, via \eqref{eq: letter identification}, into constraints that should hold for the three-point form factor:
\begin{equation}
\begin{gathered} 
\cancel{\dots d \otimes e \dots}\, ,  \\
\cancel{\dots a \otimes d \dots}\, , \quad  \cancel{\dots d \otimes a \dots}\, , 
\end{gathered}
\end{equation}
plus all dihedral images. The first of these conditions was observed in~\cite{Dixon:2020bbt,Chicherin:2020umh}, while the other conditions are new.\footnote{In fact, these additional relations were observed by the authors prior to the discovery of~\eqref{ff_amp_relation}, and they indeed hold through eight loops~\cite{ToAppearEightLoop}.}

More generally, this duality makes it possible to translate knowledge about the functional form of one of these quantities into information about the other. Most obviously, the form factor can simply be `read off' of the amplitude on the $\Delta = 0$ surface (up to $i\pi$ contributions). Conversely the form factor also provides an enormous amount of boundary data for bootstrapping the amplitude. In fact, we have checked that this information, when combined with parity, is sufficient to uniquely determine the symbol of the amplitude
through $5$ loops, and through $7$ loops when combined also with certain conditions on the final pair of entries and the behavior at the origin \cite{Basso:2020xts}.

\section{Discussion and Conclusions}

In this letter, we have identified a new and unexpected duality in planar $\mathcal{N}=4$ SYM theory between the three-point form factor of the chiral part of the stress tensor supermultiplet, and a kinematic limit of the six-point MHV amplitude. Amazingly, these quantities are related by the antipode map, which has no clear physical interpretation that we are aware of. In particular, the antipode exchanges the first and last entries of the symbol, which describe the discontinuities and derivatives of these functions, respectively. Thus, the discontinuities of the amplitude seem to be encoded in the derivatives of the form factor, and vice versa!

While we have provided evidence for this duality through seven loops, it would be interesting to find a physical derivation or even a proof of this relation, using for example the non-perturbative integrability-based descriptions of both quantities (\cite{Basso:2013vsa, Basso:2013aha,Basso:2014koa,Basso:2014nra,Basso:2014hfa,Basso:2015rta,Basso:2015uxa,Belitsky:2014sla,Belitsky:2014lta,Belitsky:2016vyq} and~\cite{Sever:2020jjx,Sever:2021nsq,Sever:2021xga}). At strong coupling both quantities can also be described via a minimal surface \cite{Alday:2007hr,Alday:2007he} and a corresponding Y-system \cite{Alday:2010vh,Maldacena:2010kp,Gao:2013dza}, and it would be interesting to see what relation \eqref{ff_amp_relation} implies for these formulations.

The three-point form factor is of particular interest due to the principle of maximal transcendentality~\cite{Kotikov:2001sc,Kotikov:2002ab,Kotikov:2004er,Kotikov:2007cy}, which states that the three-point form factor in $\mathcal{N}=4$ SYM theory provides the maximally transcendental part of the Higgs-to-three-gluon amplitude in pure Yang-Mills theory in the large-top-mass approximation \cite{Wilczek:1977zn,Shifman:1978zn,Dixon:2004za,Gehrmann:2011aa,Brandhuber:2012vm}.

It would be extremely interesting to see whether a version of the duality we present here exists for higher-point MHV amplitudes and form factors. 
Since $A_{2n}$ and $F_n$ both exhibit a $D_{n}$ dihedral symmetry, one might expect these quantities to be related also for $n>3$.
The surface $\Delta =0$ can be interpreted as `twisted forward scattering' in which the $2n$ external momenta of the amplitude, $\hat{p}_i$, are related by $\hat{p}_{i+n} = -\hat{p}_{i}$ for $i=1,2,\ldots,n$ (for $n=3$), so that there are only $n$ independent momenta, as in the form factor. This interpretation might give further clues for a generalization to higher $n$.

Moreover, it would be interesting to see whether a similar duality exists at next-to-MHV and beyond, or for other operators than the chiral part of the stress tensor supermultiplet.

\vspace{6pt}\acknowledgments
We would like to thank Benjamin Basso, Francis Brown, Claude Duhr,
Andy Liu and Cristian Vergu for stimulating discussions.  This
research was supported by the US Department of Energy under contract
DE--AC02--76SF00515.  AJM and MW were supported in part by the ERC
starting grant 757978 and grant 00015369 from  Villum Fonden. MW
was additionally supported by grant 00025445 from Villum Fonden. \"OG
is supported by the UKRI/EPSRC Stephen Hawking Fellowship EP/T016396/1.

\bibliography{../second_paper/ff}

\end{document}